\documentclass[prd,superscriptaddress,unsortedaddress,twocolumn,showpacs,floatfix]{revtex4}

\usepackage{epsfig}
\usepackage{dcolumn}
\usepackage{amsmath}

\def\Hzzz{  0.59} 
\def\HzzzERRSTAT{  0.20} 
\def\HzzzERRSYSTKTEV{  0.48} 
\def\HzzzERRSYSTEXT{  1.06} 
\def\HzzzERRTOT{  1.19}

\def\CHISQDAL{ 2805.3} 
\def\NDOFDAL{ 2765} 
 
\def\CHISQEDGE{  125.3} 
\def\NDOFEDGE{  130}

\def\NDATA{   68.3}   % Millions
\def\NMCANA{  124.9}   % Millions
\def\MCDATARATIO{    1.8} 
\def\SYSTACCID{  0.02} 
\def\SYSTSCAT{  0.05} 
\def\SYSTKPZFIT{  0.07} 
 
\def\SYSTESCALE{  0.06} 
\def\SYSTESMEAR{  0.04} 
\def\SYSTETAIL{  0.02} 
\def\SYSTPOSRES{  0.07} 
\def\SYSTNOPSF{  0.02} 
\def\SYSTMCSTAT{  0.14} 
\def\SYSTRZERO{  0.21} 
\def\SYSTRTWO{  0.04} 
\def\SYSTAPL{  0.01} 
\def\SYSTGPMZ{  0.05} 
\def\SYSTHPMZ{  0.05} 
\def\SYSTZZCUT{  0.44} 
\def\SYSTADIF{  1.03} 
\def\SYSTAZERO{  0.12} 
 
\def\SYSTSUMREC{  0.13} 
\def\SYSTSUMFIT{  0.46} 
\def\SYSTSUMEXT{  1.06} 
 
\def\VVDIF{  0.07} 
\def\VVERR{  0.09} 
 
\def\HHzzz{ -2.09} 
\def\HHzzzERRSTAT{  0.62} 
\def\HHzzzERRSYSTKTEV{  0.72} 
\def\HHzzzERRSYSTEXT{  0.28} 
\def\HHzzzERRTOT{  0.99}

\def\CHISQDALTWOD{ 2790.6} 
\def\NDOFDALTWOD{ 2764} 
\def\CHISQEDGETWOD{  126.3} 
\def\NDOFEDGETWOD{  130} 
 
\def\HHSYSTACCID{  0.03} 
\def\HHSYSTSCAT{  0.05} 
\def\HHSYSTKPZFIT{  0.07} 
 
\def\HHSYSTESCALE{  0.13} 
\def\HHSYSTESMEAR{  0.11} 
\def\HHSYSTETAIL{  0.02} 
\def\HHSYSTPOSRES{  0.10} 
\def\HHSYSTNOPSF{  0.07} 
\def\HHSYSTMCSTAT{  0.46} 
\def\HHSYSTRZERO{  0.15} 
\def\HHSYSTRTWO{  0.01} 
\def\HHSYSTAPL{  0.08} 
\def\HHSYSTGPMZ{  0.12} 
\def\HHSYSTHPMZ{  0.17} 
\def\HHSYSTZZCUT{  0.51} 
\def\HHSYSTAZERO{  0.07} 
 
\def\HHSYSTSUMREC{  0.21} 
 
\def\HHSYSTSUMEXT{  0.28} 
 
\def\AAdif{  0.215} 
\def\AAdifERRSTAT{  0.014} 
\def\AAdifERRSYSTKTEV{  0.025} 
\def\AAdifERRSYSTEXT{  0.006} 
\def\AAdifERRTOT{  0.031} 
 
\def\ADIFKTEVCERN{    1.5} 
 
\def\AASYSTACCID{  0.000} 
\def\AASYSTSCAT{  0.000} 
\def\AASYSTKPZFIT{  0.000} 
 
\def\AASYSTESCALE{  0.002} 
\def\AASYSTESMEAR{  0.002} 
\def\AASYSTETAIL{  0.000} 
\def\AASYSTPOSRES{  0.001} 
\def\AASYSTNOPSF{  0.002} 
\def\AASYSTMCSTAT{  0.011} 
\def\AASYSTRZERO{  0.001} 
\def\AASYSTRTWO{  0.000} 
\def\AASYSTAPL{  0.002} 
\def\AASYSTGPMZ{  0.004} 
\def\AASYSTHPMZ{  0.003} 
\def\AASYSTZZCUT{  0.022} 
\def\AASYSTAZERO{  0.002} 
 
\def\AASYSTSUMREC{  0.003} 
 
\def\AASYSTSUMEXT{  0.006} 
 
\def\ZZZRES{0.94}  % 3p0 mass res, MeV
\def\RDSQSIGRES{0.014}  % gauss sigma-res of R^2
\def\ZZMINRES{0.3}    % M2P0min res (MeV)
 
\def\ZZZSIDEDATA{0.21\%}  % M3p0 sideband frac
\def\ZZZSIDEMCANA{0.20\%}  
 
\def\KPZCUTDATA{0.43\%}  % KP0FIT cut-frac
\def\KPZCUTMCANA{0.47\%} 
 
\newcommand{\rec}{reconstruction}
\newcommand{\unc}{uncertainty}
\newcommand{\uncs}{uncertainties}

\newcommand{\ADIFCERN}{0.268}
\newcommand{\ADIFERRCERN}{0.017}     % ignores a0/a2 correlation
\newcommand{\ZZCUT}{0.274~{\rm GeV}/c^2}
\newcommand{\HzzzSIGN}{+}

\newcommand{\hzzz}{ h_{000} }
\newcommand{\hprime}{ h^{\prime} }

\newcommand{\hpdgsym}{ \hzzz({\rm PDG06}) }
\newcommand{\hpdgvalue}{ (-5.0 \pm 1.4)\times 10^{-3} }
\newcommand{\hPDG}{ -0.005 }
\newcommand{\adif}{a_0 - a_2}

\newcommand{\adiftwod}{{a}_0 - {a}_2}
\newcommand{\hzzztwod}{{h}_{000} }

\newcommand{\pz}{ \pi^0 }
\newcommand{\zzz}{ \pi^0\pi^0\pi^0 }
\newcommand{\pmz}{\pi^+\pi^-\pi^0}

\newcommand{\ee}{e^+e^-}
\newcommand{\pp}{\pi^+\pi^-}
\newcommand{\zz}{\pi^0\pi^0}

\newcommand{\KLpmz}{ K_{L}\to \pi^+\pi^-\pi^0 }
\newcommand{\KLzzz}{ K_L\to \pi^0\pi^0\pi^0 }
\newcommand{\KLpienu}{ K_L\to \pi^{\pm}e^{\mp}\nu }
\newcommand{\mpi}{m_{\pi^+} }
\newcommand{\mpiz}{ m_{\pi^0} }
\newcommand{\mpizz}{ m_{\pi^0\pi^0} }
\newcommand{\mpipi}{ m_{\pi^+\pi^-} }
\newcommand{\minmpizz}{ m_{\pi^0\pi^0}^{\rm min} }

\newcommand{\Kpzz}{ K^{\pm}\to \pi^{\pm}\pi^0\pi^0 }
\newcommand{\Kppp}{ K^{\pm}\to \pi^{\pm}\pi^{\mp}\pi^{\pm} }

\newcommand{\mgg}{ M_{\gamma\gamma} }

\newcommand{\unit}{ \times 10^{-3} }
\newcommand{\Mzzz}{ {\cal M}_{000}  }
\newcommand{\SQMzzz}{ \vert {\cal M}_{000} \vert^2 }
\newcommand{\ZK}{ z }

\newcommand{\RDSQ}{ R^2_D }
\newcommand{\XD}{ X_D }
\newcommand{\YD}{ Y_D }

\newcommand{\ktev}{ KTeV }

\newcommand{\Ndata}{N_{xy}^{\rm data} }
\newcommand{\Npred}{N_{xy}^{\rm pred} }
\newcommand{\Nsim}{N_{x'y'}^{\rm MC} }
\newcommand{\kpzfitchi} { \chi_{E}^2 }

\newcommand{\RGMIN}{r_{\gamma}}

\newcommand{\POLYMZZGEV}{0.281~{\rm GeV}/c^2}
\newcommand{\POLYMZZ}{0.281}
\newcommand{\RATMZZ}{ {\mathcal R}_{00}^{\rm model} }

\newcommand{\reepoe}{ Re(\epsilon^{\prime}/\epsilon) }

\newcommand{\pairchi} { \chi_{pair}^2 }

\begin{document}     % START DOCUMENT

%=======================================================================
%=======================================================================
%=======================================================================

\title{Detailed Study of the  $\KLzzz$ Dalitz Plot }
% Version for KTeV 99 papers.
%

\newcommand{\UAz}{University of Arizona, Tucson, Arizona 85721}
\newcommand{\UCLA}{University of California at Los Angeles, Los Angeles,
                    California 90095} 
\newcommand{\Campinas}{Universidade Estadual de Campinas, Campinas, 
                       Brazil 13083-970}
\newcommand{\EFI}{The Enrico Fermi Institute, The University of Chicago, 
                  Chicago, Illinois 60637}
\newcommand{\UB}{University of Colorado, Boulder, Colorado 80309}
\newcommand{\ELM}{Elmhurst College, Elmhurst, Illinois 60126}
\newcommand{\FNAL}{Fermi National Accelerator Laboratory, 
                   Batavia, Illinois 60510}
\newcommand{\Osaka}{Osaka University, Toyonaka, Osaka 560-0043 Japan} 
\newcommand{\Rice}{Rice University, Houston, Texas 77005}

\newcommand{\SaoPaulo}{Universidade de S\~ao Paulo, S\~ao Paulo, 
	Brazil 05315-970}

\newcommand{\UVa}{The Department of Physics and Institute of Nuclear and 
                  Particle Physics, University of Virginia, 
                  Charlottesville, Virginia 22901}
\newcommand{\UW}{University of Wisconsin, Madison, Wisconsin 53706}

\affiliation{\UAz}
\affiliation{\UCLA}
\affiliation{\Campinas}
\affiliation{\EFI}
\affiliation{\UB}
\affiliation{\ELM}
\affiliation{\FNAL}
\affiliation{\Osaka}
\affiliation{\Rice}
\affiliation{\SaoPaulo}
\affiliation{\UVa}
\affiliation{\UW}

\author{E.~Abouzaid}	  \affiliation{\EFI}
\author{M.~Arenton}       \affiliation{\UVa}
\author{A.R.~Barker}      \altaffiliation[Deceased.]{ } \affiliation{\UB}
\author{L.~Bellantoni}    \affiliation{\FNAL}
\author{E.~Blucher}       \affiliation{\EFI}
\author{G.J.~Bock}        \affiliation{\FNAL}
\author{E.~Cheu}          \affiliation{\UAz}
\author{R.~Coleman}       \affiliation{\FNAL}
\author{M.D.~Corcoran}    \affiliation{\Rice}
\author{B.~Cox}           \affiliation{\UVa}
\author{A.R.~Erwin}       \affiliation{\UW}
\author{C.O.~Escobar}     \affiliation{\Campinas}  %%Consider after Ke4 paper
\author{A.~Glazov}        \affiliation{\EFI}
\author{A.~Golossanov}    \affiliation{\UVa} %%Remove March 08
\author{R.A.~Gomes}       \affiliation{\Campinas}
\author{P. Gouffon}       \affiliation{\SaoPaulo}
\author{Y.B.~Hsiung}      \affiliation{\FNAL}
\author{D.A.~Jensen}      \affiliation{\FNAL}
\author{R.~Kessler}       \affiliation{\EFI}
\author{K.~Kotera}	  \affiliation{\Osaka}
\author{A.~Ledovskoy}     \affiliation{\UVa}
\author{P.L.~McBride}     \affiliation{\FNAL}

\author{E.~Monnier}
   \altaffiliation[Permanent address ]{C.P.P. Marseille/C.N.R.S., France}
   \affiliation{\EFI}  %% Doug will ping him 

\author{H.~Nguyen}       \affiliation{\FNAL}
\author{R.~Niclasen}     \affiliation{\UB}
\author{D.G.~Phillips~II} \affiliation{\UVa}
\author{E.J.~Ramberg}    \affiliation{\FNAL}
\author{R.E.~Ray}        \affiliation{\FNAL}
\author{M.~Ronquest}     \affiliation{\UVa}
\author{E.~Santos}       \affiliation{\SaoPaulo}
\author{W.~Slater}       \affiliation{\UCLA}
\author{D.~Smith}        \affiliation{\UVa}
\author{N.~Solomey}      \affiliation{\EFI}
\author{E.C.~Swallow}    \affiliation{\EFI}\affiliation{\ELM}
\author{P.A.~Toale}      \affiliation{\UB}
\author{R.~Tschirhart}   \affiliation{\FNAL}
\author{Y.W.~Wah}        \affiliation{\EFI}
\author{J.~Wang}         \affiliation{\UAz}
\author{H.B.~White}      \affiliation{\FNAL}
\author{J.~Whitmore}     \affiliation{\FNAL}
\author{M.~J.~Wilking}      \affiliation{\UB}
\author{B.~Winstein}     \affiliation{\EFI}
\author{R.~Winston}      \affiliation{\EFI}
\author{E.T.~Worcester}  \affiliation{\EFI}
\author{T.~Yamanaka}     \affiliation{\Osaka}
\author{E.~D.~Zimmerman} \affiliation{\UB}
\author{R.F.~Zukanovich} \affiliation{\SaoPaulo}

\collaboration{The KTeV Collaboration}

% -----------------------------------------------------

\date{\today}

\begin{abstract}
  Using a sample of $\NDATA$ million $\KLzzz$  decays  
  collected in 1996-1999 by the  \ktev\ (E832) 
  experiment at Fermilab,
  we present a detailed study of the $\KLzzz$ Dalitz plot density.
  We report the first observation of interference
  from $\KLpmz$ decays in which $\pp$ rescatters to $\zz$ 
  in a final-state interaction. 
  This rescattering effect is described by the 
  Cabibbo-Isidori model,  %%% \cite{cusp2},
  and it depends on the difference in pion scattering lengths 
  between the isospin $I=0$ and $I=2$ states, $\adif$.
  Using the Cabibbo-Isidori model,
  and fixing $(\adif)\mpi = \ADIFCERN \pm \ADIFERRCERN$ 
  as measured by the CERN-NA48 collaboration, %%% \cite{NA48_cusp}, 
  we present the first measurement of the $\KLzzz$ 
  quadratic slope parameter that accounts
  for the rescattering effect:
  $\hzzz = ( \HzzzSIGN\Hzzz \pm \HzzzERRSTAT_{stat} 
                            \pm \HzzzERRSYSTKTEV_{syst}
                            \pm \HzzzERRSYSTEXT_{ext}
            )\unit$, 
  where the \uncs\ are from data statistics,
  \ktev\ systematic errors, and external systematic errors.
  Fitting for both $\hzzz$ and $\adif$, we find
  $\hzzztwod = ( \HHzzz \pm \HHzzzERRSTAT_{stat} 
                    \pm \HHzzzERRSYSTKTEV_{syst}
                    \pm \HHzzzERRSYSTEXT_{ext}
            )\unit$, 
     and
  $\mpi(\adiftwod) = \AAdif \pm \AAdifERRSTAT_{stat} 
                   \pm \AAdifERRSYSTKTEV_{syst}
                  \pm \AAdifERRSYSTEXT_{ext}$~;
  our value for $\adif$ is consistent with that from NA48.
\end{abstract}

\maketitle

\tableofcontents

% #######################
%
%       START 
%
% #######################

% #####################################################
  \section{Introduction}
  \label{sec:intro}
% #####################################################

%=======================================================================
%
%  NOTE: 
%    x = v/sqrt(3);  v = x * sqrt(3)
%    y = u
%
%=======================================================================

The amplitude for the $K_L\to \zzz$ decay includes
contributions from two sources. 
The first source is from intrinsic dynamics that represent
a $K_L$ decaying directly into the $\zzz$ final state.
The second contribution is from the decay $K_L\to\pmz$ 
followed by a rescattering, $\pp\to \zz$.
The amplitudes from these two contributions result in a
small $(\sim 1\%)$ interference pattern in the Dalitz plot density.

The Dalitz plot density corresponding to the intrinsic $K\to 3\pi$
decay amplitude is approximately described by~\cite{pdg06}
\begin{equation}
  \vert {\cal M}^{int}(\XD,\YD) \vert^2 \propto 1 + g\YD + \sqrt{3}j\XD 
               + h\YD^2 + 3k\XD^2~,
\end{equation}
where
\begin{eqnarray}
  \XD &  \equiv & (s_1 - s_2)/(\sqrt{3}\mpi^2)  \\
  \YD &  \equiv & (s_3 - s_0)/\mpi^2  \\      
  s_i &  \equiv & (P_K - P_i)^2,~~i=1,2,3   \\
  s_0 &  \equiv & (s_1 + s_2 + s_3)/3 ~,
\end{eqnarray}
and $P_K$ and $P_i$ are the four-momenta of the parent kaon 
and the three pions. 
The linear $g,j$ parameters and the quadratic $h,k$ parameters
are determined experimentally.
For the specific case of $\KLzzz$ decays,
the linear terms vanish, $h=3k$ and the intrinsic Dalitz 
plot density reduces to~\cite{Messner74,Devlin79}
\begin{equation}
   \vert\Mzzz^{\rm int}\vert^2    \propto  1 + \hzzz\RDSQ ~,
   \label{eq:hdef}
\end{equation}
where 
\begin{eqnarray}
  \RDSQ  &  = & \XD^2 + \YD^2  \label{eq:rdsqdef}
         \\
        &  = & \frac{4}{\mpi^4} \left[
                   s_0^2 - (s_1 s_2 + s_1 s_3 + s_2 s_3)/3
                           \right]~,
    \label{eq:rsqdef1}
\end{eqnarray}
and $\hzzz$ is the quadratic slope parameter.
With $|\hzzz| < 10^{-2}$ and
$\RDSQ < 2.5$ from kinematic constraints, 
the variation of  $\SQMzzz$ over the entire Dalitz plot is 
less than 2\%.

% E731 RSQ-res = 0.02 * 5 = 0.10 ?
% NA48 RSQ-res = 0.03
% KTEV RSQ-res = 0.014

There are two previous measurements of $\hzzz$.
The first reported measurement, 
from Fermilab experiment E731~\cite{dal3pi0_e731},
is based on 5~million recorded $\KLzzz$ decays with an $\RDSQ$-resolution
$\sigma(\RDSQ) \sim 0.1$ determined from Monte Carlo simulations.
CERN experiment NA48~\cite{dal3pi0_na48} used
nearly 15~million decays with $\sigma(\RDSQ) \simeq 0.03$.
The average of these two results is
$\hpdgsym = \hpdgvalue$~\cite{pdg06}.
Here we report a more precise result from \ktev\ based on 
$\NDATA$~million decays with $\sigma(\RDSQ) \simeq \RDSQSIGRES$.
Compared with the previous measurements of $\hzzz$,
a major difference in the \ktev\ analysis is that 
we take into account the contribution
from $\KLpmz$ decays in which $\pp \to \zz$ in a 
final-state interaction.

A full treatment of rescattering in $K\to 3\pi$ decays,
including higher order loop corrections, 
is given by Cabibbo and Isidori \cite{cusp2}.
This model, referred to hereafter as CI3PI,
describes a ``cusp'' in the region where the 
minimum $\zz$ mass is very near $2\mpi$:
a cusp refers to a localized region of the Dalitz plot
where the density changes very rapidly.
A precisely measured shape of this cusp can be used to  
measure the difference in pion scattering lengths between
the isospin $I=0$ and $I=2$ states, $\adif$.
In 2006, the CERN-NA48 collaboration reported the first observation 
of a cusp in $\Kpzz$ decays.
The interference effect is from the decay $\Kppp$ 
followed by rescattering: $\pp\to\zz$.
They reported $(\adif)\mpi = \ADIFCERN \pm \ADIFERRCERN$ \cite{NA48_cusp},
in excellent agreement with the prediction of 
Chiral Perturbation Theory \cite{CGL2000,CGL2001}.

\begin{figure*}
  \centering
  \epsfig{file=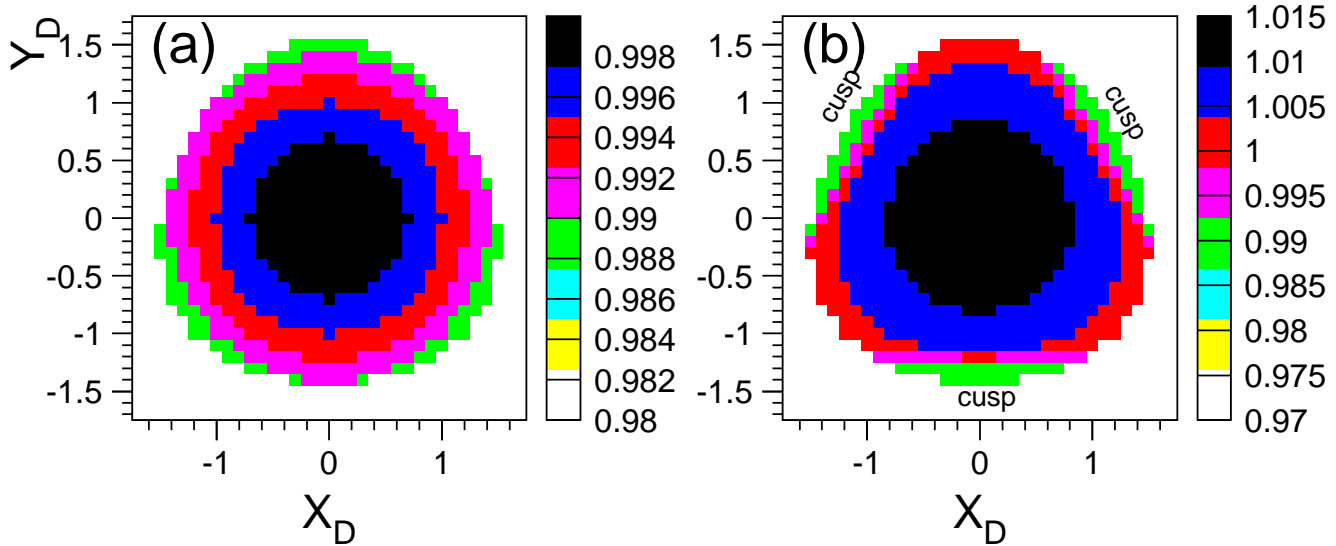,width=\linewidth}
  \caption{
     Expected deviation from $\KLzzz$ phase-space based on
     (a) $\hzzz = \hPDG$ and no contribution from rescattering, and
     (b) $\KLpmz$ with rescattering as calculated by 
       Cabibbo and Isidori \cite{cusp2}, with $\hzzz = 0$.    
     The intensity scales are slightly different to   
     better illustrate the relative shapes.
          }
  \label{fig:xydal_model}
\end{figure*}

Compared to $\Kpzz$ decays,
a much smaller cusp is expected in $\KLzzz$ decays,
and here we report the first such observation as part of our 
measurement of the quadratic slope parameter.
The expected distortion of the $\KLzzz$ Dalitz plot is shown in 
Fig.~\ref{fig:xydal_model}a using $\hzzz = \hPDG$ and no 
contribution from rescattering,
and in Fig.~\ref{fig:xydal_model}b using $\hzzz = 0$ and
CI3PI \cite{cusp2} to model rescattering from $\KLpmz$.

The effects of rescattering and a negative value of $\hzzz$
both result in the Dalitz plot density dropping slowly as
$\RDSQ$ increases. The maximum variation is only a few percent.
The main feature that separates these two effects is that
while the quadratic slope parameter results in a smooth linear
function of $\RDSQ$, the rescattering from $\KLpmz$ results
in a much sharper fall-off near some of the Dalitz plot edges
(see  ``cusp'' labels in Fig.~\ref{fig:xydal_model}b). 
Also note that the Dalitz plot density in 
Fig.~\ref{fig:xydal_model}a is azimuthally symmetric,
while the density in  Fig.~\ref{fig:xydal_model}b 
is azimuthally asymmetric.
This cusp will become more apparent when we
examine the minimum $\zz$ mass in \S~\ref{sec:cusp}.

The outline of this report is as follows.
The \ktev\ detector and simulation are described
in \S~\ref{sec:detector}-\ref{sec:MC}.
The \rec\ of $\KLzzz$ decays and the determination
of the Dalitz plot variables ($\XD,\YD$) are presented
in \S~\ref{sec:recon}.
\S~\ref{sec:hzzz_fit} describes the fitting technique used
to extract the quadratic slope parameter ($\hzzz$) and 
the difference in scattering lengths ($\adif$).
Systematic uncertainties are described in \S~\ref{sec:syst},
and \S~\ref{sec:hresult} presents results for $\hzzz$ with
$\adif$ fixed to the value measured by the NA48 collaboration.
In \S~\ref{sec:adif}, both the quadratic slope parameter
and the difference in scattering lengths are determined
simultaneously in a two-parameter fit of the 
$\KLzzz$ phase space.

% #####################################################
  \section{Experimental Apparatus}
  \label{sec:detector}
% #####################################################

The \ktev\ detector has been described in detail 
elsewhere~\cite{ktev03_reepoe,prd_br}.
Here we give a brief description of the essential detector components.
An 800~GeV proton beam incident on a beryllium-oxide (primary) 
target produces neutral kaons along with other charged and 
neutral particles. Sweeping magnets remove charged particles
from the beamline.
Beryllium absorbers 20 meters downstream of the target 
attenuate the beam in a manner that increases the kaon-to-neutron ratio.
A collimation system results in two parallel
neutral beams beginning 90~meters from the primary target;
each beam consists of roughly equal numbers of kaons and neutrons.
The fiducial decay region is 121-158 meters from the target,
and the vacuum region extends from 20-159~meters.
A regenerator, designed to produce $K_S$ decays for the $\reepoe$ measurement,
alternates between the two beams.
The other neutral beam is called the vacuum ($K_L$) beam.
Only $\KLzzz$ decays from the vacuum beam are used in this analysis.

\begin{figure}[hb]
  \centering
  \epsfig{file=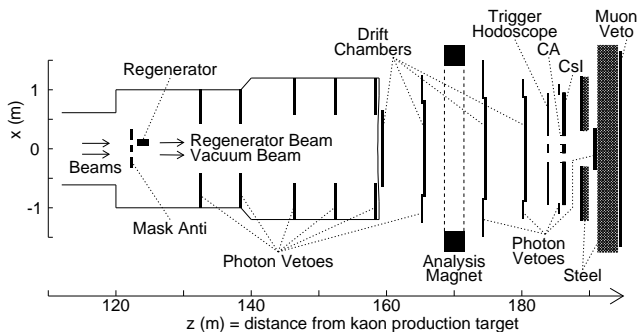,width=\linewidth}
  \caption{
    Plan view of the \ktev\ (E832) detector.
    The evacuated decay volume ends with a thin vacuum window at
    $Z = 159$~m.  Only decays from the vacuum beam are used
    to measure $\hzzz$.
           }
  \label{fig:detector}
\end{figure}

\begin{figure}
  \centering
  \epsfig{file=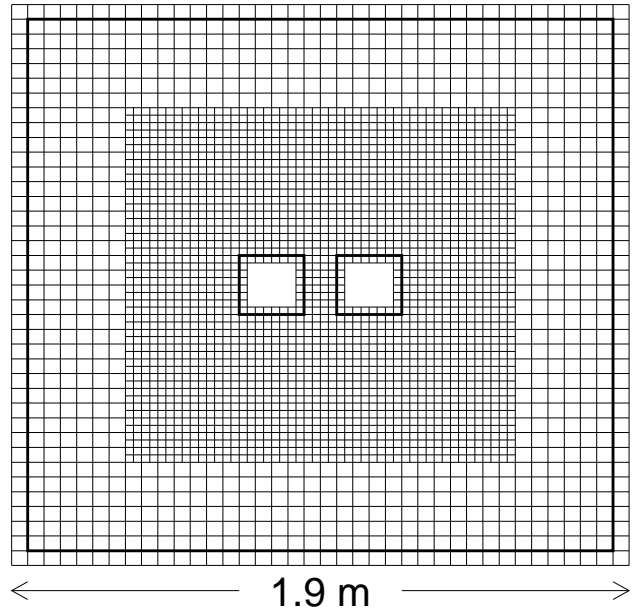,width=\linewidth}
  \caption{
       Layout of CsI calorimeter. 
       The two neutral beams go through the
       $15\times 15~{\rm cm}^2$ beam holes (into page) 
       shown by the two inner squares.  
       The fiducial cut, indicated by the dark lines,
       rejects $\KLzzz$ decays in which any photon hits
       a crystal adjacent to a beam hole or at the outer boundary.
          }
  \label{fig:csi}
\end{figure}

The \ktev\ detector (Fig.~\ref{fig:detector})
is located downstream of the decay region.
The main element used in the analysis is 
an electromagnetic calorimeter made of 3100 pure 
cesium iodide (CsI) crystals (Fig.~\ref{fig:csi}).
For photons and electrons,
the energy resolution is better than 1\%
and the position resolution is about 1~mm.
The CsI calorimeter has two holes to
allow the neutral beams to pass through without interacting.

A spectrometer consisting of four drift chambers,
two upstream and two downstream of a dipole magnet, 
measures the momentum of charged particles;
the resolution is 
$\sigma_p/p \simeq [1.7 \oplus (p/14) ] \times 10^{-3}$,
where $p$ is the track momentum in GeV/$c$.
Bags filled with helium are placed between the drift chambers
and inside the magnet, replacing about 25~meters of air.
A scintillator ``trigger'' hodoscope just upstream of the
CsI is used to trigger on decays with charged particles
in the final state.
The \ktev\ beamline has very little material upstream of the
CsI calorimeter, thereby reducing the impact of
external photon conversions ($\gamma X \to X e^+e^-$).
The total amount of material is 0.043 radiation lengths,
about half of which is in the trigger hodoscope.
Eight photon-veto detectors along the decay region and spectrometer
reject events with escaping particles.

An electronic trigger for $\KLzzz$ decays requires 
at least 25~GeV total energy deposit in the CsI calorimeter,
as well as six isolated clusters with energy above 1~GeV.
For $\KLzzz$ decays that satisfy the energy and vertex requirements
(\S~\ref{sec:recon}),
approximately 9\% of these decays satisfy the six-cluster trigger,
and 20\% of the six-cluster events were recorded for analysis.
The combined six-cluster data from three run periods 
(1996, 1997, 1999)
has nearly 400 million recorded events.
$\KLzzz$ decays are recorded for use as a high-statistics crosscheck
on the Monte Carlo (see below) determination of the acceptance
in the $\reepoe$ analysis \cite{ktev03_reepoe}.
This  sample is ideal to study the Dalitz density.

% ---------------------------------
  \section{Monte Carlo Simulation}
  \label{sec:MC}
% ---------------------------------

A Monte Carlo simulation (MC) is used to determine the expected
$\KLzzz$ Dalitz plot density that would be observed without 
the contribution from rescattering and with $\hzzz = 0$; 
i.e, pure phase space.
The $\KLzzz$ dynamics are determined from deviations between 
the observed Dalitz density (Fig.~\ref{fig:xydal_data})
and that from the phase-space MC.
The simulated phase-space density accounts for 
detector geometry, detector response, trigger, 
and selection requirements in the analysis.

The simulation of $\KLzzz$ decays begins by selecting the kaon 
momentum from a distribution measured with $K \to \pp$ decays.
Each simulated kaon undergoes scattering
in the beryllium absorbers near the target,
and kaons that hit the edge of any collimator
are either scattered or absorbed. 
For kaons that scatter from a collimator edge,
the $K_L$-$K_S$ mixture has been determined from a study of 
$\KLpmz$ and $K^0\to\pp$ decays.
After generating a kaon trajectory downstream of the collimator,
each photon from $\KLzzz\to 6\gamma$ is traced through the detector, 
allowing for external ($\gamma\to\ee$) conversions.
The secondary electron-positron pairs are traced through the detector
and include the effects of multiple scattering, 
energy loss from ionization, and bremsstrahlung.
The effects of accidental activity are 
included by overlaying events
from a trigger that recorded random activity in the detector 
that is proportional to the instantaneous intensity of the proton beam.

For photons and electrons that hit the CsI calorimeter,
the energy response is taken from a 
shower library generated with {\sc geant} \cite{geant}:
each library entry contains the energy deposits in a $13\times 13$
grid of crystals centered on the crystal struck by the incident particle.
The shower library is binned in incident energy, 
position within a crystal, and angle.

For both data and MC,
the energy calibration for the CsI is performed with
momentum-analyzed electrons from $\KLpienu$ decays. 
To match CsI energy resolutions for data and MC,
an additional 0.3\% fluctuation is added to 
the MC energy response.
The data also show a low-side response tail that is not present 
in the MC, and is probably due to photo-nuclear interactions
in the CsI calorimeter.
As explained in Appendix~B of \cite{prd_br},
this energy-loss tail has been accurately measured with
electrons from $\KLpienu$ decays, and this tail is empirically
modeled in the simulation with the assumption that the 
energy-loss tail is the same for photons and electrons.
Losses up to 40\% of the incident photon/electron energy 
are included in the model.

The CsI position resolution is measured with precise electron 
trajectories in $\KLpienu$ decays. The position
resolution for the MC is found to be nearly 10\% worse than for data,
requiring that the MC cluster positions be ``un-smeared'' to match
the data resolution.  The un-smearing is done for each simulated photon
cluster by moving the reconstructed position closer to the
true (generated) position in the CsI calorimeter.
The position un-smearing fraction is $0.07 + 0.0016E_{\gamma}$,
where $E_{\gamma}$ is the photon energy in GeV.

Nearly five billion $\KLzzz$ decays were generated by our
Monte Carlo simulation. More than 90\% of the generated decays 
are rejected by the geometric requirement that all six photons 
hit the CsI calorimeter; about 2/3 of these six-cluster  events
are rejected by the selection requirements described below
in \S~\ref{sec:recon}.
The resulting sample of $\NMCANA$ million reconstructed 
$\KLzzz$ decays corresponds to 
$\MCDATARATIO \times$ the data statistics.

% #####################################################
  \section{Reconstruction of $\KLzzz$ Decays}
  \label{sec:recon}
% #####################################################

The \rec\ of $\KLzzz \to 6\gamma$ is based on measured energies and 
positions of photons that hit the CsI calorimeter.
Exactly six clusters, each with  a transverse profile consistent 
with a photon, are required.
The cluster positions must be separated by at least $7.5$~cm,
and each cluster energy must be greater than  $3$~GeV. 
For the two nearest photon clusters in the CsI calorimeter, 
we require that the minimum-to-maximum photon energy ratio 
is greater than 20\%;
this requirement eliminates the most extreme cases
of overlapping clusters in which an energetic photon
lands very close to a photon of much lower energy.
The fiducial volume is defined by 
cluster positions measured in the calorimeter;
we reject events in which any reconstructed photon position
is in a crystal adjacent to a beam-hole or 
in the outermost layer of crystals (Fig.~\ref{fig:csi}).

To remove events in which the kaon has scattered
in the collimator or regenerator, we define
the center-of-energy of the six photon clusters  to be
\begin{equation}
  x_{ce} = \sum_{i} x_i E_i/\sum_i E_i ~~~~~~~~
  y_{ce} = \sum_{i} y_i E_i/\sum_i E_i
\end{equation}
where $x_i,y_i$ are the measured photon positions in the
CsI calorimeter, $E_i$ are the measured photon energies,
and the index $i=1,6$.
The coordinate $x_{ce},y_{ce}$ is the point where the kaon would 
have intercepted the plane of the calorimeter if it had not decayed.
The size of each beam at the CsI calorimeter 
is about $10 \times 10~{\rm cm}^2$;
the center-of-energy, measured with $\sim 1$~millimeter resolution,
is required to lie within an $11 \times 11~{\rm cm}^2$ 
square centered on the kaon beam.

Photons are paired to reconstruct three
neutral pions consistent with a single decay vertex. 
There are 15 possible photon pairings for
a $\KLzzz$ decay. To select the best $\zzz$ pairing,
we introduce a ``pairing-$\chi^2$'' variable ($\pairchi$),  
which quantifies the consistency of the three $\pi^0$ vertices.
To ensure a reliable \rec\ of the Dalitz variables,
we require that the smallest of the 15 $\pairchi$ values 
is less than 10 (the mean $\pairchi$ is 3), and also
that the second smallest $\pairchi$ value is greater than 30.
The  location of the kaon decay vertex ($\ZK$) is determined from a 
weighted average of the  $\pz$ vertices.

\begin{figure}
  \centering
  \epsfig{file=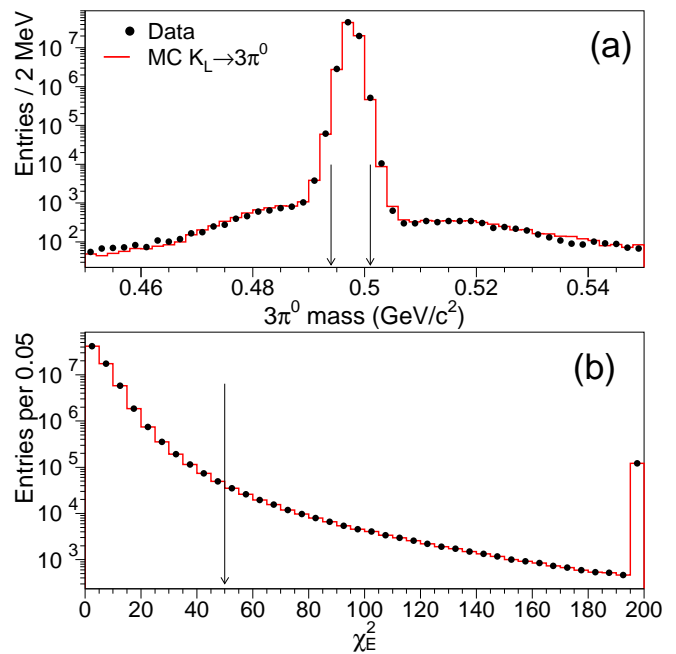,width=\linewidth}
  \caption{
     (a) Invariant $\zzz$ mass with all selection requirements
     except for the $\zzz$-mass and $\kpzfitchi$.
     The $\zzz$ mass resolution (from Gaussian fit) is
     $\ZZZRES~{\rm MeV}/c^2$.
     (b) shows $\kpzfitchi$ distribution with all other selection 
         requirements; last bin includes all events with
         $\kpzfitchi > 200$.
     Dots are data and the histogram is MC.     
     Vertical arrows show the selection requirements.
          }
  \label{fig:neutmass}
\end{figure}

The main kinematic requirement is that the invariant mass of
the $\zzz$ final state
be between $0.494$ and $0.501$~GeV/$c^2$, or nearly $\pm 4\sigma$.
Figure~\ref{fig:neutmass}a shows the $\zzz$-mass distribution
for data and the MC.
The mass side-bands are due to $\KLzzz$ decays in which
the wrong photon pairing is found in the \rec.
The fraction of reconstructed events outside the invariant mass
cut is $\ZZZSIDEDATA$ for data and $\ZZZSIDEMCANA$ for the MC,
confirming that the MC provides an excellent description of the data.

Additional selection requirements are that the energy-sum of the 
six photon clusters lie between 40 and 160~GeV, and that the 
reconstructed decay vertex
is within 121--158~meters from the primary target.
To prevent an accidental cluster from faking a photon, 
we use the energy-vs-time profiles recorded by the CsI calorimeter. 
For each photon candidate, the CsI cluster energy deposited in a 
19~nanosecond window before the event must be consistent with pedestal.
To limit the effect of external photon conversions in the
detector material ($\gamma X \to Xe^+e^-$),
we allow no more than one hit in the scintillator hodoscope
that lies 2~meters upstream of the CsI calorimeter.

To improve the resolution of the Dalitz plot parameters
($\XD,\YD,\RDSQ$), the cluster energies
are adjusted for each event by imposing kinematic constraints
to minimize
\begin{equation}
   \kpzfitchi = \sum_{i=1}^6 \frac{(E_i-E_i^{fit})^2}{\sigma_i^2}~,
    \label{eq:kp0fit}
\end{equation}
where $E_i$ are the reconstructed cluster energies,
$\sigma_i$ are the energy resolutions,
and $E_i^{fit}$ are the six fitted cluster energies.
The impact of cluster position resolution on the Dalitz parameters
is much smaller than that of the energy resolution, 
and therefore the cluster positions are fixed in the minimization.
The four kinematic constraints are $M_{6\gamma} = M_K$
and $\mgg = \mpiz$ for each of the three neutral pions.
With four constraints and six unknowns in Eq.~\ref{eq:kp0fit},
the minimization has two degrees of freedom.
Events with $\kpzfitchi < 50$ are selected for the analysis.
The minimization of $\kpzfitchi$ improves 
the $\RDSQ$-resolution  from 0.070  to  $\RDSQSIGRES$.
Fig.~\ref{fig:neutmass}b shows a data-MC comparison
of the $\kpzfitchi$ distribution.
The fraction of events removed by the $\kpzfitchi$ cut is
$\KPZCUTDATA$ for data and $\KPZCUTMCANA$ for the MC;
this slight disagreement will be addressed in the evaluation
of systematic \uncs.

\begin{figure}
  \centering
  \epsfig{file=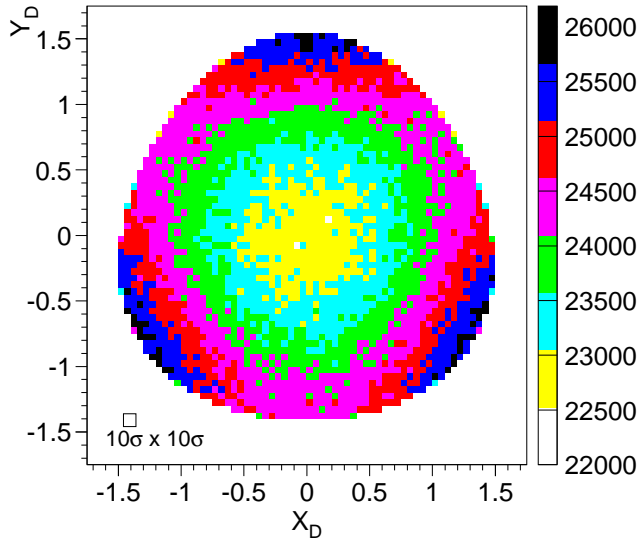,width=\linewidth}
  \caption{
     Dalitz plot density, $\YD$ vs. $\XD$,
     for $\NDATA$ million $\KLzzz$ decays
     in the \ktev\ data sample after all
     selection requirements.
     The color-scale at right shows the number
     of events in each  $0.05 \times 0.05$  pixel.
     The \rec\ resolution on $\XD$ and $\YD$  is 
     $\sigma \sim 0.01$ as determined by the MC;
     the box in the lower-left corner shows
     $10\sigma \times 10\sigma$ for illustration.
          }
  \label{fig:xydal_data}
\end{figure}

After all \rec\ and selection requirements
there are $\NDATA$~million $\KLzzz$ events.
The two-dimensional Dalitz plot distribution for this sample is shown
in Fig.~\ref{fig:xydal_data} with no acceptance correction,
and projections onto $\RDSQ$ and the minimum $\zz$ mass
are shown in Fig.~\ref{fig:m2p0rsq}a,b.
The $\sim 5$\% variation across the Dalitz plot
is mainly from the detector acceptance;
this variation is nearly an order of magnitude 
larger than the variations from rescattering
(CI3PI) and the quadratic slope parameter.
Also note that the uncorrected phase space distribution
has a minimum at the center of the Dalitz plot,
while the expected effects from physics (Fig.~\ref{fig:xydal_model})
result in a maximum at the center.
An accurate $\KLzzz$ simulation 
is therefore critical to this measurement.

% #####################################################
  \section{Fit for $\hzzz$ and $\adif$ }
  \label{sec:hzzz_fit}
% #####################################################

\begin{figure*}
  \centering
  \epsfig{file=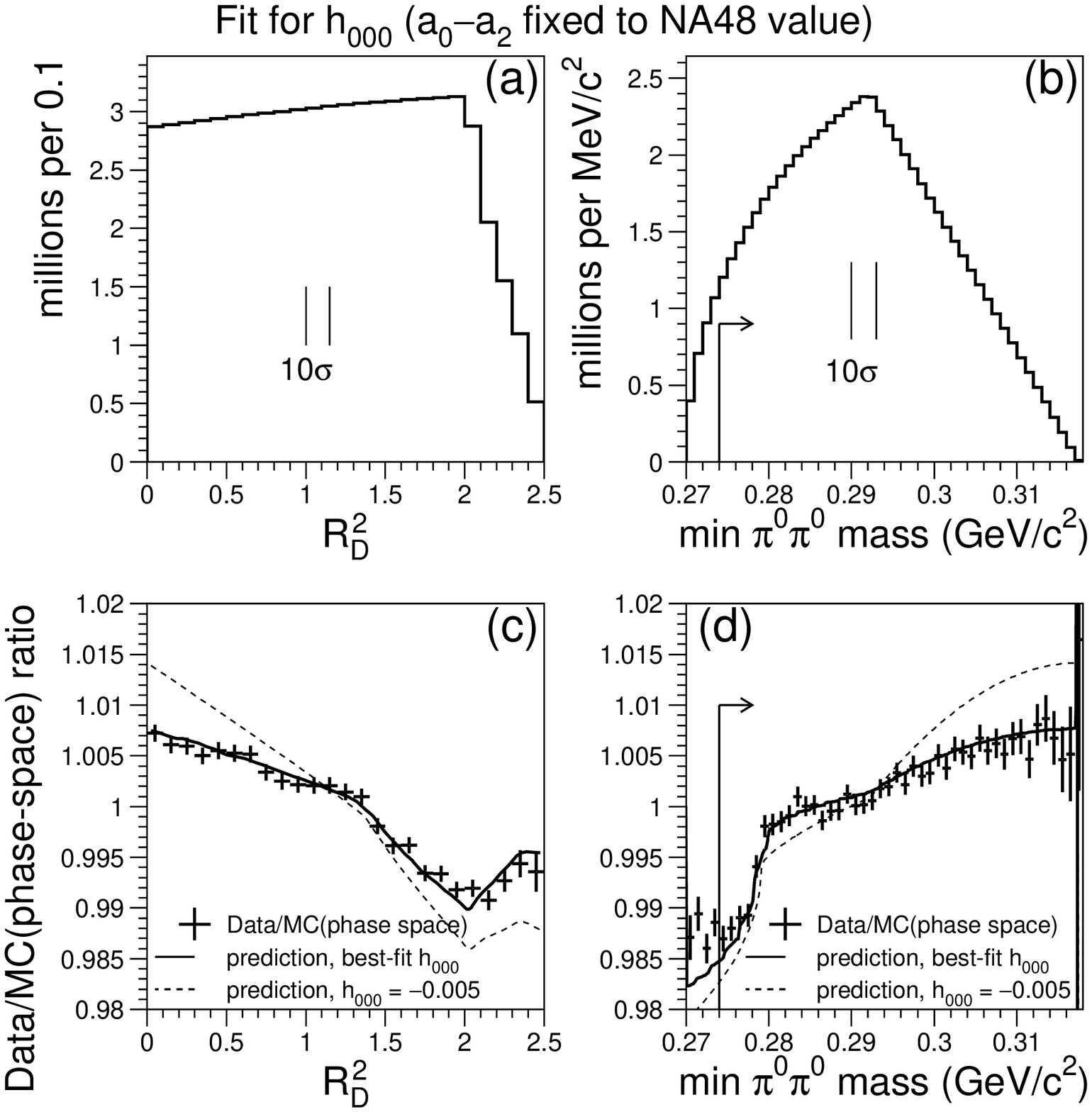,width=\linewidth}
  \caption{  
     For the $\NDATA$~million $\KLzzz$ in the \ktev\ sample,
     projected Dalitz distributions are shown for
     (a) $\RDSQ$ and (b) $\minmpizz$.
     The average \rec\ resolution determined by the simulation is
     $\sigma(\RDSQ) \sim \RDSQSIGRES$ and 
     $\sigma({\rm min}~\mpizz) \sim \ZZMINRES$~MeV$/c^2$:
     these resolutions are indicated by a
     $10\sigma$ marker on each plot.     
     The data/MC(phase-space) ratio is shown as a function of 
     (c) $\RDSQ$ and (d) $\minmpizz$ (points with error bars).
     The solid curve is the prediction from our best fit $\hzzz$.
     The dashed curve is the prediction using 
     $\hpdgsym = \hpdgvalue$.   
     The arrow in (d) shows the selection requirement
     $\minmpizz > \ZZCUT$.
     Note that previous analyses \cite{dal3pi0_e731,dal3pi0_na48} 
     ignored rescattering and excluded $\RDSQ>1.9$ ; 
     the corresponding data/MC ratio
     was assumed to be a straight line with slope of $-0.005$.
          }   
  \label{fig:m2p0rsq}
\end{figure*}

In the previous measurements of the quadratic slope parameter $\hzzz$
\cite{dal3pi0_e731,dal3pi0_na48},
the $\RDSQ$-distribution for data (Fig.~\ref{fig:m2p0rsq}a) 
was compared to the $\RDSQ$ distribution from a simulated $\KLzzz$ 
sample with $\hzzz = 0$.
Normalizing the MC sample to have the same data statistics at
$\RDSQ=0$,
the data/MC ratio was fit to a linear function,
$1 + \hprime\RDSQ$, where $\hprime$ is the fitted slope.
The region $\RDSQ > 1.9$ was excluded because this
region is more sensitive to energy nonlinearities 
and resolution.
The $\KLzzz$ slope parameter was then assumed to be $\hzzz = \hprime$.
Applying the same procedure to our \ktev\ data yields a result
consistent with the CERN-NA48 result \cite{dal3pi0_na48},
but with a very poor fit probability.

In light of new information about rescattering from $\KLpmz$,
we fit our two-dimensional Dalitz plot to the CI3PI model \cite{cusp2}.
With the exception of $\hzzz$ and $\adif$, the CI3PI model parameters 
have been measured or calculated theoretically,
and these parameters are listed in Table~\ref{tb:modelpar}.
In this section the fitting technique is described within the
framework of a single-parameter fit for $\hzzz$ with 
the value of $\adif$ fixed by an external measurement (from NA48).
However, this fitting technique works the same way when
both $\hzzz$ and $\adif$ are floated in the fit.

\begin{table}[hb]
  \centering
  \caption{
    Parameters and their values used in the CI3PI model.
    $K_{3\pi}$ refers to $\KLpmz$,
    subscripts 0 and 2 refer to isospin $I=0$ and $I=2$.
       }
  \medskip
\begin{ruledtabular}
\begin{tabular}{lc}
% ------------------------------------
 parameter   &   value \\
\hline % ------------------------------------------------
$K_{3\pi}$ linear slope ($g_{+-0}$)    & $0.678 \pm 0.008 $ \cite{pdg06} \\
$K_{3\pi}$ quadratic slope ($h_{+-0}$) & $0.076 \pm 0.006 $ \cite{pdg06} \\
$a_0\mpi$ at $\pp$ thresh       &  $0.216 \pm 0.013$ \cite{BNL865_a0}      \\
$(\adif)\mpi$ at $\pp$ thresh   &  $\ADIFCERN \pm \ADIFERRCERN$ 
                                       \cite{NA48_cusp}  \\
Effective ranges ($r_0, r_2$)  & $1.25\pm 0.04$, $1.81 \pm 0.05$ \cite{cusp2}   \\
$A_L^+/A_L^0$                   &   $0.28 \pm 0.03$  \cite{cusp2}      \\
Isospin breaking parameter ($\epsilon$)  &   0.065  \cite{NA48_cusp}  \\
\end{tabular}
\end{ruledtabular}
    \label{tb:modelpar}
\end{table}

We first define $\Ndata$ to be the number of
events reconstructed in a $0.05\times 0.05$
Dalitz pixel (Fig.~\ref{fig:xydal_data})
denoted by integers $x$ and $y$.
The model prediction for the number of events in 
each Dalitz pixel, $\Npred$, is given by
\begin{eqnarray}
   \Npred = ~~~~~~~~~~~~~~~~~~   \nonumber \\
   {\cal N} \sum_{^{x'=x-2,x+2} _{y'=y-2,y+2}}
         \vert\Mzzz(x',y')\vert^2 \Nsim {\rm PSF}(x'-x,y'-y)~.
   \label{eq:Npred}
\end{eqnarray}
The quantities appearing in the prediction function are
explained as follows.
${\cal N}$ is an overall normalization factor such that the  
total phase-space density integrals on each side of 
Eq.~\ref{eq:Npred} are the same.
$\Mzzz(x',y')$ is the matrix element at the center of 
pixel $x',y'$, as calculated from the CI3PI model
and the floated value of $\hzzz$.
The remaining quantities are based on $\KLzzz$ MC generated
with $\hzzz=0$ and no rescattering: i.e., flat phase space.
$\Nsim$ is the number of $\KLzzz$ events {\it generated} 
in pixel $x',y'$ that pass all selection criteria;
note that $\Nsim$ is not the number of MC events reconstructed
in pixel $x',y'$.
${\rm PSF}(x'-x,y'-y)$ is the ``pixel-spread-function,'' 
computed from MC, which gives the fraction of events generated
in pixel $x',y'$ that are reconstructed in pixel $x,y$.
In each of the 2956 Dalitz pixels with data, 
the ${\rm PSF}$ is computed on a $5\times 5$ grid around the pixel.
The pixel size corresponds to about 
$5\sigma \times 5\sigma$ in terms of the \rec\ resolution
of $\XD$ and $\YD$.
On average, 70\% of the MC events are reconstructed
in the same Dalitz pixel as the generation pixel;
99.96\% of the MC events are reconstructed
within a $3\times 3$ pixel grid centered on
the generation pixel.

The data are fit with {\sc minuit} to minimize the $\chi^2$-function
\begin{equation}
   \chi^2 = \sum_{x,y} \left[
        (\Ndata - \Npred) / \sigma_{xy}^{\rm pred}
    \right]^2 ~,
    \label{eq:chi2}
\end{equation}
where $\Npred$ is the prediction function in Eq.~\ref{eq:Npred},
and $\Ndata$ is the number of reconstructed $\KLzzz$
decays in pixel $x,y$. The statistical \unc\ is 
\begin{equation}
   (\sigma_{xy}^{\rm pred})^2 = 
      \Npred + (\Npred)^2/N_{xy}^{\rm MC} 
\end{equation}
where $N_{xy}^{\rm MC}$ is the number of MC events reconstructed
in pixel $x,y$. The two terms above represent the statistical 
\unc\ on the data and MC, respectively.

In the fitting procedure we make an additional
selection requirement that among the three possible
$\zz$ pairings, the minimum $\zz$ mass, ``$\minmpizz$,''
is greater than $\ZZCUT$. 
This requirement removes 3~million (4.5\%) $\KLzzz$ decays 
from the data sample,
and it is applied because of a slight data-model discrepancy
that is discussed in 
\S~\ref{subsec:syst_fit} and \S~\ref{subsec:crosschecks}.

The quality of the fit is illustrated by the $\chi^2$.
With $\hzzz$ as a fit parameter and $\adif$ fixed,
$\chi^2/{\rm dof} = \CHISQDAL / \NDOFDAL$ for all of the
pixels, and  $\chi^2_{\rm edge}/{\rm dof}  =  \CHISQEDGE / \NDOFEDGE$
for the subset of edge-pixels that overlap the
Dalitz boundary.
The sensitivity of $\chi^2_{\rm edge}$ is illustrated by
fitting the data without the kinematically-constrained
energy adjustments (Eq.~\ref{eq:kp0fit}):
in this case, $\chi^2_{\rm edge}$ increases by 60.
Fitting for both $\hzzz$ and $\adif$, the corresponding
$\chi^2$ and $\chi^2_{\rm edge}$ are very similar.
The results of these fits are presented in
\S~\ref{sec:hresult}-\ref{sec:adif}.

% #####################################################
  \section{Observation of Interference from $\KLpmz$ with Rescattering}
  \label{sec:cusp}
% #####################################################

While the cusp from rescattering is clearly visible
in the CERN-NA48 distribution of 
$\mpizz$ from $\Kpzz$ decays (see Fig.~2 of \cite{NA48_cusp}),
there is no such evidence in our raw distribution of $\minmpizz$
from $\KLzzz$ decays (Fig.~\ref{fig:m2p0rsq}b).
The rescattering effect in $\KLzzz$ decays becomes apparent
only when the data are divided by the corresponding MC distribution
generated with pure phase-space: 
i.e, $\hzzz=0$ and no rescattering
from $\KLpmz$ decays. 
These data/MC(phase-space) ratios are shown as data points with 
errors in Figs.~\ref{fig:m2p0rsq}c,d.
A cusp is clearly visible in the Dalitz region
$\RDSQ \sim 2$ and $\minmpizz \sim 2\mpi =0.28~{\rm GeV}$.
The rescattering process $\pp\to\zz$ changes from a virtual process
($\mpipi < 2\mpi$) resulting in destructive interference,
to a real process ($\mpipi > 2\mpi$)
resulting in constructive interference.

We use the fit results (\S~\ref{sec:hresult}) to compute a 
prediction for the data/MC(phase-space) ratio as a function 
of $\RDSQ$ and $\minmpizz$;
these predictions are shown as solid curves in 
Figs.~\ref{fig:m2p0rsq}c,d.
The predictions agree well with our measured data/MC(phase-space)
distributions, except for the discrepancy in the region
defined by $\minmpizz < \ZZCUT$ 
(first four bins of Fig.~\ref{fig:m2p0rsq}d).
The dashed curves show the prediction using the CI3PI model
and $\hzzz$ replaced with the current PDG value, $\hzzz = \hPDG$;
these curves clearly do not match the \ktev\ distributions.
To easily reproduce the KTeV prediction, we have parametrized
the solid curve in Fig.~\ref{fig:m2p0rsq}d as a polynomial
of the form:
\begin{equation}
   \RATMZZ(\minmpizz) = \sum_{n=0}^3 m_i\times(\minmpizz - \POLYMZZ)^n~,
   \label{eq:polym2p0}
\end{equation}
where $\minmpizz$ is the minimum $\zz$ mass (GeV$/c^2$),
and the coefficients ($m_{n=0,3}$) are given in 
Table~\ref{tb:polym2p0}.  
The root-mean-square precision of this parametrization 
is 0.023\%, and the largest deviation of the
parametrization is 0.06\%.

\begin{table}[hb]
  \centering
  \caption{
     Polynomial coefficients (Eq.~\ref{eq:polym2p0})  
     for the parametrization of the solid curve in 
     Fig.~\ref{fig:m2p0rsq}d for 
     $\hzzz = \HzzzSIGN\Hzzz$ and $\mpi(\adif) = \ADIFCERN$.
     Note that the coefficients depend on the value of $\minmpizz$.
       }
  \medskip
\begin{ruledtabular}
\begin{tabular}{l | cccc}
% ------------------------------------
 valid range of   &         &         &        &        \\
 min $\zz$ mass   &  $m_0$  &  $m_1$  & $m_2$  & $m_3$  \\
\hline % ------------------------------------------------
$\minmpizz < \POLYMZZGEV$ 
   &  0.999937   & 3.34994   & 165.229     & 0 \\
$\minmpizz > \POLYMZZGEV$ 
   &  0.998851 & 0.121152  & 16.5534 &  $-372.656$  \\
\end{tabular}
\end{ruledtabular}
  \label{tb:polym2p0}
\end{table}

% #####################################################
  \section{ Systematic Uncertainties}
  \label{sec:syst}
% #####################################################

Systematic \uncs\ are broken into three
categories: detector \& reconstruction, fitting, 
and external parameters.
Within the framework of a single-parameter fit for $\hzzz$
these categories are discussed in the subsections below,
and the systematic \uncs\ on $\hzzz$ are 
summarized in Table~\ref{tb:syst}.
The\ktev\ detector and analysis introduces a systematic \unc\ 
of $\HzzzERRSYSTKTEV\unit$ on $\hzzz$.
Uncertainties in external parameters, particularly $\adif$,
lead to a much larger \unc\ of $\HzzzERRSYSTEXT\unit$
on $\hzzz$.
For the two-parameter fit 
($\hzzztwod$ and $\adiftwod$; see \S~\ref{sec:adif}), 
the systematic \uncs\ are evaluated in the same manner,
and these \uncs\ are summarized in Table~\ref{tb:syst2D}.
Note that when a systematic variation results in a shift that
is comparable to the the statistical \unc, 
we make an effort to justify an \unc\ that is smaller
than the systematic variation;
when the corresponding shift is much smaller than the
statistical \unc, there is no need to justify a smaller \unc.

\begin{table}[hb]
  \centering
  \caption{
   Systematic \uncs\ on $\hzzz$.
   For each external parameter $\cal X$, the sign ($+$ or $-$) 
   is indicated for the partial derivative,
   $\partial\hzzz/\partial{\cal X}$, so that our $\hzzz$ result can
   be updated when an external parameter is updated.
       }
  \medskip
\begin{ruledtabular}
\begin{tabular}{lc}
% ------------------------------------
 source of   &  \multicolumn{1}{c}{\unc\ on $\hzzz$}   \\
 \unc\       &  \multicolumn{1}{c}{$(\unit)$ }         \\
\hline % ------------------------------------------------
DETECTOR \& RECON   &             \\
~~~kaon scattering           &  $\SYSTSCAT$    \\
~~~accidentals               &  $\SYSTACCID$   \\
~~~photon energy scale       &  $\SYSTESCALE$      \\
~~~energy resolution         &  $\SYSTESMEAR$      \\
~~~low-side energy tail      &  $\SYSTETAIL$       \\
~~~position resolution       &  $\SYSTPOSRES$      \\
~~~$\kpzfitchi$-cut          &  $\SYSTKPZFIT$      \\
(sub-total)               & $(\SYSTSUMREC)$ \\
\hline
FITTING                          &              \\
~~~MC statistics                 & $\SYSTMCSTAT$   \\
~~~Ignore PSF for $\Npred$       & $\SYSTNOPSF$    \\
~~~remove $\minmpizz$ cut    & $\SYSTZZCUT$   \\
(sub-total)                  & ($\SYSTSUMFIT$) \\
\hline
\ktev\  TOTAL     &   $\HzzzERRSYSTKTEV$  \\
\hline
                  &   \\
\hline
EXTERNAL                 &           \\
~~~$(\adif)\mpi$         &  $(+)$ $\SYSTADIF$                \\
~~~$a_0\mpi$             &  $(-)$ $\SYSTAZERO$               \\
~~~$r_0,~r_2$            &  $(+)$ $\SYSTRZERO$,~$(+)$ $\SYSTRTWO$   \\
~~~$A_L^+/A_L^0$         &  $(+)$ $\SYSTAPL$                \\
~~~$g_{+-0},~h_{+-0}$    &  $(-)$ $\SYSTGPMZ$,~$(-)$ $\SYSTHPMZ$    \\
(sub-total)    & ($\SYSTSUMEXT$) \\
\end{tabular}
\end{ruledtabular}
  \label{tb:syst}
\end{table}

% -------------------------------------
  \subsection{Detector \& Reconstruction}
  \label{subsec:syst_det}
% -------------------------------------

Systematic \uncs\ on $\hzzz$ are mainly from effects
that bias the reconstructed Dalitz variables,
$\XD$ and $\YD$, in a manner that is not accounted for
in the simulation.

\bigskip\noindent {\bf Kaon Scattering} \\
Recall that beryllium absorbers were placed 20 meters downstream
of the primary target in order to increase the kaon-to-neutron ratio.
Scattering in these absorbers affects the kaon trajectory, 
and hence the reconstructed Dalitz variables.
If absorber-scattering is turned off in the MC,
the resulting value of $\hzzz$ changes by $0.5\unit$.
Based on studies of kaon trajectories with
$K^0\to\pp$ decays in the vacuum beam, 
we assign a systematic \unc\ on $\hzzz$ equal to 10\% 
of the change when scattering is turned off in the simulation:
$\SYSTSCAT\unit$.

\bigskip\noindent{\bf Accidental Activity} \\
Energy deposits from accidental activity in the CsI calorimeter 
can modify the reconstructed photon energies.
In the reconstruction, events are rejected if any of the six
photon clusters has accidental activity within a
19~nanosecond window prior to the start-time of the event.
Removing this cut increases the level of accidental activity,
and changes $\hzzz$ by $\SYSTACCID\unit$; we include this 
difference as a systematic \unc.

\bigskip\noindent{\bf Photon Energy Scale} \\
The photon energy scale is determined in the $\reepoe$ analysis
by comparing the data and MC vertex distributions for 
$K^0\to\zz$ decays downstream of the regenerator. 
These decays are mainly due to the $K_S$-component of the
neutral kaon. 
The active veto system rejects decays inside the regenerator,
resulting in a rapidly rising distribution just
downstream of the regenerator.
The data-MC vertex comparison has a discrepancy that is slightly
dependent on kaon energy, and the magnitude  
of the discrepancy no more than 3~cm; 
this data-MC shift in the vertex corresponds
to an energy-scale discrepancy of up to $\sim 0.05$\%.
An energy scale correction is empirically derived to
remove this small discrepancy in $K^0\to\zz$ decays,
and this ``$\zz$'' correction is applied to photon energies 
in the $\KLzzz$ Dalitz analysis.
As a systematic test, the Dalitz analysis is performed
with no energy scale correction: $\hzzz$ changes by 
$\SYSTESCALE\unit$ and is included as a systematic error.

\bigskip\noindent{\bf Photon Energy Resolution} \\
The simulated energy resolution is adjusted by about 0.3\%
to match the energy resolution for electrons from
$\KLpienu$ decays.
The resulting photon energy resolution is well simulated,
as illustrated by the excellent data-MC agreement
in the $\zzz$-mass distribution (Fig.~\ref{fig:neutmass}a).
As a systematic test, we increase the simulated resolution
by an additional 0.3\%:
the change in $\hzzz$ is $\SYSTESMEAR\unit$,
and is included as a systematic \unc.

\bigskip\noindent{\bf Low-Side Energy Tail} \\
The effects of photo-nuclear interactions and wrapping material
in the CsI calorimeter can result in photon energies measured
well below a few-sigma fluctuation in the expected photostatistics.
As described in \S~\ref{sec:MC}, this non-Gaussian tail
has been measured using electrons from $\KLpienu$ decays,
and modeled in the simulation. Based on the data-MC agreement
in the low-side $E/p$ tail for electrons, we assign a 20\% \unc\
on our understanding of this effect.
As an illustration, note that the Gaussian energy resolution 
(0.8\%) predicts that 0.02\% of the photons will be reconstructed 
with an energy that is at least 3\% below the true value; 
the effect of the non-Gaussian tail is that 0.8\% of the 
reconstructed photon energies are at least 3\% low.
As a systematic test, we remove simulated decays in which
any photon loses more than 3\% of its energy due to
this non-Gaussian process.
This test rejects $6\times 0.8\% \sim 5\%$ of the 
generated $\KLzzz$ decays. 
After applying selection requirements, the MC sample is reduced by 2\%,
which is smaller than the reduction for generated decays.
Using this test-MC sample, the change in $\hzzz$ is $0.1\unit$
compared to using the nominal MC;
as explained above, we include 20\% of this change, $\SYSTETAIL\unit$,
as a systematic \unc\ on $\hzzz$.

\bigskip\noindent{\bf Photon Position Resolution} \\
Turning off the ``un-smearing'' (\S~\ref{sec:MC}) 
of the MC photon positions results in a change of
$0.3\unit$ in $\hzzz$. Based on the data-MC agreement
in the electron position resolution from $\KLpienu$ decays,
we take 20\% of this change,
$\SYSTPOSRES\unit$, as a systematic \unc.

\bigskip\noindent{\bf $\kpzfitchi$ Cut} \\
The determination of the Dalitz variables is performed
using adjusted photon energies, where the adjustment
is done for each $\KLzzz$ decay by minimizing
the ``energy-$\kpzfitchi$'' in Eq.~\ref{eq:kp0fit}.
The selection requirement is $\kpzfitchi < 50$.
As a systematic test, this cut is relaxed to $\kpzfitchi < 1000$;
the change in $\hzzz$ is $\SYSTKPZFIT\unit$,
and is included as a systematic \unc.

% -------------------------------------
  \subsection{Fitting}
  \label{subsec:syst_fit}
% -------------------------------------

\bigskip\noindent{\bf MC Statistics} \\
The  simulated sample consists of $\NMCANA$ million $\KLzzz$ decays
that satisfy the selection requirements
($\MCDATARATIO \times$ the data statistics).
This sample results in a MC-statistics \unc\ 
of $\SYSTMCSTAT\unit$ on $\hzzz$.

\bigskip\noindent{\bf Pixel Migration} \\
The reconstructed pixel location in the Dalitz plot ($\XD,\YD$)
can be different than the true pixel location
This pixel migration is accounted for by using the 
pixel-spread-function (PSF)
in Eq.~\ref{eq:Npred} to predict the number of reconstructed 
$\KLzzz$ decays in each pixel.
As a systematic test, we ignore pixel migration by setting
${\rm PSF}(x'-x,y'-y) = \delta(x'-x,y'-y)$ and replacing
$\Nsim$ with the number of events reconstructed in each pixel;
the change in $\hzzz$, $\SYSTNOPSF\unit$,
is included as a systematic \unc.

\bigskip\noindent{\bf Data-Model Discrepancy} \\
As shown in Fig.~\ref{fig:m2p0rsq}d,
the Dalitz region defined by $\minmpizz < \ZZCUT$  
shows a data-model discrepancy, and this region
is therefore excluded from the nominal fit.
Including this region in the fit changes $\hzzz$ by $\SYSTZZCUT\unit$,
and we include this difference as a systematic \unc.
Additional discussion on this discrepancy is given 
in \S~\ref{subsec:crosschecks}.

% -------------------------------------
  \subsection{External Parameters}
  \label{subsec:syst_ext}
% -------------------------------------

The CI3PI model depends on several parameters listed in
Table~\ref{tb:modelpar}. The \uncs\ in these parameters
have been propagated through the $\hzzz$ fit.
The net $\hzzz$ \unc\ from these external parameters
is $\SYSTSUMEXT\unit$. This \unc\ is almost entirely
due to the \unc\ in the difference in scattering lengths,
$\adif$.

In Table~\ref{tb:syst}, we have also included the sign
of each partial derivative so that our $\hzzz$ result
can be updated when an external parameter is updated.
For example, $\partial\hzzz/\partial 
(a_0\mpi) = -\SYSTAZERO/0.013$,
where the numerator and denominator are from 
Tables~\ref{tb:syst} and \ref{tb:modelpar}, respectively.

% #####################################################
  \section{Result for $\hzzz$ with Fixed $\adif$}
  \label{sec:hresult}
% #####################################################

Here we fix $\mpi(\adif) = \ADIFCERN$ as measured by 
NA48 \cite{NA48_cusp}), and determine $\hzzz$.
The result from minimizing the $\chi^2$ 
in Eq.~\ref{eq:chi2} is
\begin{eqnarray}
   \hzzz & = & (\HzzzSIGN\Hzzz \pm \HzzzERRSTAT_{stat})\unit   \\
   \chi^2/{\rm dof}  & = & \CHISQDAL / \NDOFDAL  ~~({\rm all~pixels})  
     \label{eq:allchi2} \\
   \chi^2/{\rm dof}  & = & \CHISQEDGE / \NDOFEDGE ~~ ({\rm edge~pixels})~.
     \label{eq:edgechi2}
\end{eqnarray}
where the statistical \unc\ is from
$\NDATA$ million decays in the data sample.
To check our modeling near the Dalitz boundary,
the $\chi^2$ is shown for the subset of ``edge pixels'' 
that overlap the Dalitz boundary.

Including the systematic \unc,
the final result for the quadratic slope parameter is
\begin{eqnarray}
   \hzzz & = & (\HzzzSIGN\Hzzz \pm \HzzzERRSTAT_{stat} 
                      \pm \HzzzERRSYSTKTEV_{syst} 
                      \pm \HzzzERRSYSTEXT_{ext} 
               )\unit  \nonumber \\
         &   &     \label{eq:hzzz_result}    \\
         & = & (\HzzzSIGN\Hzzz \pm \HzzzERRTOT )\unit 
\end{eqnarray}
where the \uncs\ are from data statistics,
\ktev\ systematic errors, and external systematics errors.

% #####################################################
  \subsection{Crosschecks on $\hzzz$}
  \label{subsec:crosschecks}
% #####################################################

% Asym = (EGmax-EGmin)/SUM
% Cut-ranges:  0.0, 0.53, 0.63, 0.69, 0.75, 1.0
%
% Here we define r = EGmin/EGmax = (1-A)/(1+A)
% In terms of r, the cut-ranges are
%  
% 0.31-1  0.23-.31  0.18-0.23  0.14-.18  0-.14

\begin{figure}
  \centering
  \epsfig{file=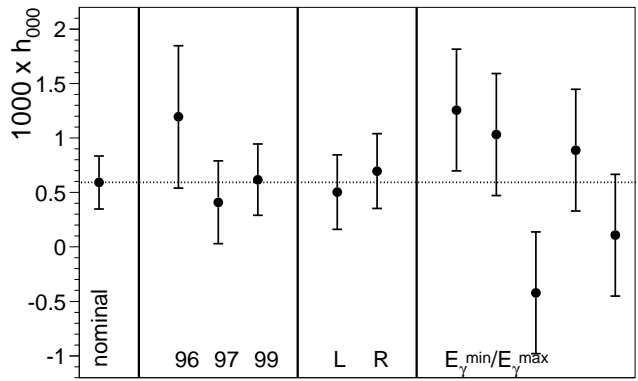,width=\linewidth}
  \caption{  
    Crosscheck measurements of $\hzzz$:
    data-taking years (96,97,99), left and right vacuum beams (L,R),
    and min/max photon-energy ratio 
    ($E_{\gamma}^{\rm min} / E_{\gamma}^{\rm max}$) 
    as discussed in the text.
    Measurements within each category (between vertical lines)
    are statistically independent.
    Error bars reflect the statistical \uncs\ from 
    the data and MC samples.
          }
  \label{fig:crosschecks}
\end{figure}

Some crosschecks on the result for $\hzzz$ 
are shown in Fig.~\ref{fig:crosschecks}.
The separate measurements for each year are consistent,
as well as the separate measurements from each vacuum beam.
The last crosscheck involves the asymmetry between the
minimum and maximum photon energy, which could expose
potential problems related to non-linearities in the 
photon energy measurement. 
The ratio between the minimum and maximum photon energies,
$\RGMIN \equiv E_{\gamma}^{\rm min} / E_{\gamma}^{\rm max}$,
is used to define five sub-samples with roughly equal statistics:
$\RGMIN = 
\{0,0.14\},\{0.14,0.18\},\{0.18,0.23\},\{0.23,0.31\},\{0.31,1\}$.
The five independent measurements of $\hzzz$ are consistent.

Concerning the data-model discrepancy in the Dalitz plot region
$\minmpizz < \ZZCUT$ (Fig.~\ref{fig:m2p0rsq}d),
we have performed many checks
to investigate if the problem is related to our analysis.
For example, the MC energy resolution was degraded by an 
additional 0.8\%, an extreme change that is nearly three
times larger than the standard 0.3\% smearing:
the corresponding change in $\hzzz$ is $1.6$
times the statistical \unc\ ($\sigma^h_{stat}$), 
but the data-model discrepancy remains unchanged.
In another test, an extreme energy nonlinearity of 
0.3\% per 100~GeV is introduced into the simulated 
energy measurements;
$\hzzz$ changes by $0.5\sigma^h_{stat}$, and the data-model
discrepancy is again unchanged.
These highly exaggerated tests suggest that the \ktev\
energy reconstruction is not responsible for the data-model
discrepancy. 
We have also checked that the data-model discrepancy is
unchanged for the following tests:
vary best $\pairchi$ cut between 4 and 100 (nominal cut is 10),
remove requirement that the second smallest $\pairchi$ 
value is greater than 30, 
allow no hits and up to six hits in the scintillator hodoscope
(to check photon conversions),
allow photons to hit a CsI crystal adjacent to the beam holes
(Fig.~\ref{fig:csi}),
remove requirement on CsI cluster energy deposited before event
(increases effect from accidentals),
vary cut on $\kpzfitchi$ from $<10$ to no cut
(Fig.~\ref{fig:neutmass}b),
remove simulated decays in which any photon loses more than
3\% of its energy in the CsI 
(see systematic test ``Low-Side Energy Tail'' in \S~\ref{subsec:syst_det}),
use reconstructed CsI photon energies instead of adjusted energies
based on kinematic constraints.

Photon conversions in detector material result in $\ee$ pairs
that are reconstructed as a single photon. A scintillator hodoscope
just upstream of the CsI calorimeter tags such $\ee$ pairs.
The standard analysis allows up to one hit in this hodoscope.
As a systematic test, we compare results with 
(i) no requirement on hodoscope hits, and with 
(ii) a requirement that there are no hits in the hodoscope.
For these two samples, there is a 15\% difference in the number
of reconstructed  $\KLzzz$ decays, and the difference in
$\hzzz$ is $(\VVDIF \pm \VVERR)\unit$.

As a final crosscheck, the analysis is repeated using the
reconstructed photon energies instead of the adjusted energies 
based on kinematic constraints from the $K_L$ and $\pz$ masses
(see $\kpzfitchi$ in Eq.~\ref{eq:kp0fit}).
Using unconstrained Dalitz variables,
the resulting value of $\hzzz$ changes by 
$1.2\sigma^h_{stat}$ compared to the nominal result.
However, compared to the nominal result in 
Eqs.~\ref{eq:allchi2}-\ref{eq:edgechi2},
the overall fit-$\chi^2$ increases by 120, and the
fit-$\chi^2$ for the edge pixels increases by nearly 60.
This increase in $\chi^2$ indicates that the resolution is 
not modeled as well for the unconstrained Dalitz variables,
and it illustrates the importance of the kinematic constraints.

% #####################################################
  \section{ Measurement of $\adif$ and $\hzzz$ with $\KLzzz$ Decays}
  \label{sec:adif}
% #####################################################

\def\RHOHZZZADIF{+0.939}

\begin{figure*}
  \centering
  \epsfig{file=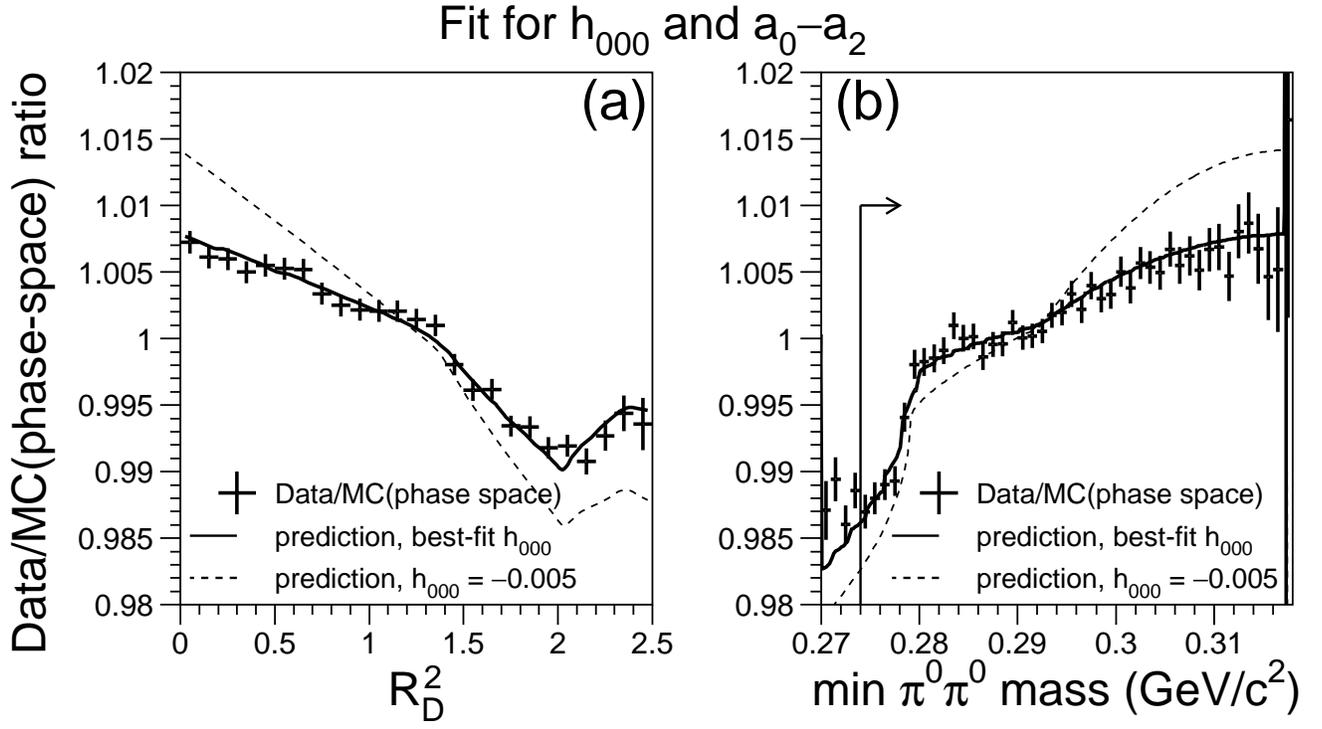,width=\linewidth}
  \caption{  
     Same as Fig.~\ref{fig:m2p0rsq}c-d, except
     $\adif$ is floated in the fit instead of
     fixed to the NA48 value.
          }   
  \label{fig:m2p0rsq_2d}
\end{figure*}

Here we use $\KLzzz$ decays to measure both
the quadratic slope parameter and the difference
in pion scattering lengths.
The fit procedure is described in \S~\ref{sec:hzzz_fit},
but now we float $\adif$ instead of fixing it to the
value measured by NA48 \cite{NA48_cusp}.
Fitting our data for both $\hzzztwod$ and $\adiftwod$
in a two-parameter fit, we find
\begin{eqnarray}
   {\hzzztwod} & = &  \nonumber \\ 
       (\HHzzz \pm \HHzzzERRSTAT_{stat} 
             & \pm & \HHzzzERRSYSTKTEV_{syst} 
                \pm \HHzzzERRSYSTEXT_{ext} 
       )\unit   
   \label{eq:HHzzz}       \\
         & = & (\HHzzz \pm \HHzzzERRTOT )\unit 
\end{eqnarray}
\begin{eqnarray}
 \mpi(\adiftwod) & = & \nonumber \\
   \AAdif \pm  \AAdifERRSTAT_{stat} 
           &  \pm & \AAdifERRSYSTKTEV_{syst} 
             \pm \AAdifERRSYSTEXT_{ext} 
          \\
         & = &  \AAdif \pm \AAdifERRTOT  \\
         &   &    \nonumber     \\
   \rho_{ha}    & = & \RHOHZZZADIF  \\
   \chi^2/{\rm dof}  & = & \CHISQDALTWOD / \NDOFDALTWOD  
        ~~({\rm all~pixels})  \\
   \chi^2/{\rm dof}  & = & \CHISQEDGETWOD / \NDOFEDGETWOD 
        ~~ ({\rm edge~pixels})~.
\end{eqnarray}
The \uncs\ are from data statistics,
\ktev\ systematic errors, and external systematic errors.
The systematic \uncs\ are evaluated in the same manner
as for the one-parameter fit for $\hzzz$ (\S~\ref{sec:syst}):
these \uncs\ are summarized in Table~\ref{tb:syst2D}.
The data-model comparisons are shown in  Fig.~\ref{fig:m2p0rsq_2d}.

Compared to the fit in which $\adif$ is fixed
(Eq.~\ref{eq:hzzz_result}),
the statistical \unc\ on $\hzzz$ is more than $\times 3$ larger
but the total \unc\ is slightly smaller.
The reason for the smaller $\hzzz$-\unc\ when $\adif$ is floated is
related to the nonlinear dependence of the correlation between
$\hzzz$ and $\adif$. 
When $\adif=\ADIFCERN$ is fixed , $d\hzzz/d(\adif) \simeq 0.06$.
For our best-fit value of $\adif=\AAdif$,  
$d\hzzz/d(\adif) \simeq 0.04$ and hence $\hzzz$ is less
sensitive to variations in $\adif$.
The asymmetry between $+1\sigma$ and $-1\sigma$ variations
is about 10\%, so we simply averaged the $\pm 1\sigma$ variations
and quote symmetric errors.

% -----------------------------------------
  \subsection{Comparisons of Results}
  \label{subsec:discuss}
% -----------------------------------------

We begin by comparing the $\hzzz$ result for the
two different fits.
Compared to the one-parameter fit where $\adif$ is fixed,
the statistical \unc\ on $\hzzztwod$ from the two-parameter fit
(Eq.~\ref{eq:HHzzz}) is about $\times 3$ larger
and the \ktev\ systematic \unc\ is $\times 1.5$ larger.
The systematic \unc\ increases by less than the statistical \unc\ 
because the largest source of \unc\ (cut on $\minmpizz$)
is similar in both the one- and two-parameter fits.
While the $\hzzz$ measurement errors are much larger for
the two-parameter fit, the external \unc\ is $\times 4$
smaller than the external \unc\ for the one-parameter fit.
The large difference in the external \uncs\ is driven by the
large correlation ($\rho_{ha} =  \RHOHZZZADIF$) between
$\hzzz$ and $\adif$. 
The overall \unc\ on $\hzzz$ is nearly the same 
for the one- and two-parameter fits;
after accounting for the different sources of \unc\ in each fit,
the significance on the different values of $\hzzz$
($\HzzzSIGN\Hzzz$ vs. $\HHzzz$) is estimated to be $2\sigma$.

Next we compare our $\adif$ result to the NA48 analysis 
based on $\Kpzz$ decays where they reported
$\adif = \ADIFCERN \pm \ADIFERRCERN$.
The \ktev\ statistical \unc\ on $\adiftwod$ is about 40\% 
larger \footnote{In reference \cite{NA48_cusp}, it is not clear if the NA48
                statistical \uncs\ include or exclude MC statistics.}
even though our $\KLzzz$ sample is more than twice as large
as their (NA48) $\Kpzz$ sample;
the larger statistical \unc\ from $\KLzzz$ decays is due to the much
smaller rescattering effect compared to $\Kpzz$ decays.
Our overall \unc\ on $\adiftwod$ is nearly $\times 2$ larger than
that obtained by NA48.
The \ktev\ and NA48 results on $\adiftwod$ are consistent
at the level of $\ADIFKTEVCERN\sigma$.
Our result is also compatible with the DIRAC result based
on measuring the lifetime of the $\pp$ atom:
$\vert \adif\vert = 0.264^{+0.033}_{-0.020}$ \cite{DIRAC2005}.

\begin{table}[hb]
  \centering
  \caption{
   Systematic \uncs\ on $\hzzztwod$ and $\mpi(\adiftwod)$.
   For each external parameter $\cal X$, the sign ($+$ or $-$) 
   is indicated for the partial derivative,
   $\partial\hzzz/\partial{\cal X}$, so that our results can
   be updated if an external parameter is updated.
       }
  \medskip
\begin{ruledtabular}
\begin{tabular}{lcc}
% ------------------------------------
 source of   &  \multicolumn{2}{c}{\unc\ on }      \\
 \unc\       &  $10^3\times\hzzztwod$   &   $\mpi(\adiftwod)$   \\
\hline % ------------------------------------------------
DETECTOR \& RECON            &                  &    \\
~~~kaon scattering           &  $\HHSYSTSCAT$   & $\AASYSTSCAT$    \\
~~~accidentals               &  $\HHSYSTACCID$  & $\AASYSTACCID$   \\
~~~photon energy scale       &  $\HHSYSTESCALE$ & $\AASYSTESCALE$  \\
~~~energy resolution         &  $\HHSYSTESMEAR$ & $\AASYSTESMEAR$  \\
~~~low-side energy tail      &  $\HHSYSTETAIL$  & $\AASYSTETAIL$   \\
~~~position resolution       &  $\HHSYSTPOSRES$ & $\AASYSTPOSRES$  \\
~~~$\kpzfitchi$-cut          &  $\HHSYSTKPZFIT$ & $\AASYSTKPZFIT$  \\
(sub-total)            & $(\HHSYSTSUMREC)$      &  $(\AASYSTSUMREC)$ \\
\hline
FITTING                         &                  &  \\
~~~MC statistics                & $\HHSYSTMCSTAT$  & $\AASYSTMCSTAT$   \\
~~~Ignore PSF for $\Npred$      & $\HHSYSTNOPSF$   & $\AASYSTNOPSF$    \\
~~~remove $\minmpizz$ cut       & $\HHSYSTZZCUT$   & $\AASYSTZZCUT$   \\
\hline
\ktev\  TOTAL  &  $\HHzzzERRSYSTKTEV$  & $\AAdifERRSYSTKTEV$  \\
\hline
                  &  &  \\
\hline
EXTERNAL                     &                 &  \\
~~~$a_0\mpi$        & $(-)$ $\HHSYSTAZERO$  & $(-)$ $\AASYSTAZERO$   \\
~~~$r_0$            & $(+)$ $\HHSYSTRZERO$  & $(+)$ $\AASYSTRZERO$   \\
~~~$r_2$            & $(-)$ $\HHSYSTRTWO$   & $(-)$ $\AASYSTRTWO$    \\
~~~$A_L^+/A_L^0$    & $(-)$ $\HHSYSTAPL$    & $(-)$ $\AASYSTAPL$     \\
~~~$g_{+-0}$        & $(+)$ $\HHSYSTGPMZ$   & $(+)$ $\AASYSTGPMZ$    \\
~~~$h_{+-0}$        & $(-)$ $\HHSYSTHPMZ$   & $(-)$ $\AASYSTHPMZ$    \\
(sub-total)         & ($\HHSYSTSUMEXT$)     & ($\AASYSTSUMEXT$)      \\
\end{tabular}
\end{ruledtabular}
  \label{tb:syst2D}
\end{table}

% #####################################################
  \section{Conclusion}
  \label{sec:conclude}
% #####################################################

We have made the first observation of interference
between the  $\KLzzz$ decay amplitude, and the amplitude for
$\KLpmz$ with the final-state rescattering process
$\pp\to\zz$.
When comparing our data to a Monte Carlo sample
of $\KLzzz$ decays generated with pure phase-space,
we see a cusp in the data/MC distribution-ratio
of minimum $\zz$ mass.
This cusp is not visible in the data distribution 
(Fig.~\ref{fig:m2p0rsq}b); 
rather, it is visible only in the data/MC ratio
(Fig.~\ref{fig:m2p0rsq}d).

Using the CI3PI model \cite{cusp2} to account for rescattering,
and fixing $\adif$ to the value measured with 
$\Kpzz$ decays \cite{NA48_cusp},
we have measured the quadratic slope parameter,
  $\hzzz = ( \HzzzSIGN\Hzzz \pm \HzzzERRTOT )\unit $, 
where the largest source of \unc\ is from the \unc\
on $\adif$.
This result is consistent with zero, 
and it disagrees with the average of previous measurements
that did not account for rescattering.
The CI3PI model describes the data well for most of the
$\KLzzz$ phase space, but there is a notable $0.3\%$ discrepancy
in the region where the minimum $\zz$ mass is less than
$\ZZCUT$. We have excluded this discrepant region from our
nominal fits, but have included this region to evaluate
systematic uncertainties. 
To investigate the possibility that the data-model 
discrepancy is from our analysis, we have made extreme
variations in the simulation of the photon energy scale
and resolution (\S~\ref{subsec:crosschecks}) and found
that such drastic changes have no impact on the discrepancy.
We have not been able to numerically verify the calculation
of the model, but for future comparisons we have left
a convenient parametrization 
(Eq.~\ref{eq:polym2p0} and Table~\ref{tb:polym2p0}).

We have repeated our phase space analysis by floating
$\adiftwod$ rather than fixing it to the value reported by NA48.
Detailed results are presented in \S~\ref{sec:adif}.
Our value of $\adiftwod$ is consistent with that
found by NA48, but with an \unc\ that is nearly 
twice as large.

%%%%%%%%%%%%%%%%%%%%%%%%%%%%%%%%%%%%%%%%%%%%%%%%%%%%%%%%%%%%
%                     Acknowledgment                       %
%%%%%%%%%%%%%%%%%%%%%%%%%%%%%%%%%%%%%%%%%%%%%%%%%%%%%%%%%%%%

We gratefully acknowledge the support and effort of the Fermilab
staff and the technical staffs of the participating institutions for
their vital contributions.  This work was supported in part by the U.S.
Department of Energy, The National Science Foundation, The Ministry of
Education and Science of Japan,
Fundacao de Amparo a Pesquisa do Estado de Sao Paulo-FAPESP,
Conselho Nacional de Desenvolvimento Cientifico e Tecnologico-CNPq and
CAPES-Ministerio Educao.
We also wish to thank Gino Isidori for helpful discussions on 
implementing the CI3PI model for $\KLzzz$ decays.

% ==============================================================
%
%                   BIBLIOGRAPHY
  \bibliography{dal3pi0}  
  \end{document}